%% file: ms.tex
\pdfoutput=1
\documentclass[10pt, conference, compsocconf]{IEEEtran}
\ifCLASSINFOpdf
   \usepackage[pdftex]{graphicx}
\else
\fi
\usepackage[cmex10]{amsmath}
\usepackage{amssymb}
\usepackage{blindtext}
\usepackage[hyphens]{url}
\usepackage{esvect}
\usepackage{pifont}

\begin{document}

\title{NEXUS: Using Geo-fencing Services without revealing your Location}


\author{\IEEEauthorblockN{Michael Guldner, Torsten Spieldenner*, Ren\'e Schubotz}
\IEEEauthorblockA{German Research Center for Artificial Intelligence (DFKI)\\
(* Saarbr\"ucken Graduate School of Computer Science)\\
66123 Saarbr\"ucken, Germany\\
Email: \{firstname.lastname\}@dfki.de}
}


%


\IEEEoverridecommandlockouts
\IEEEpubid{\begin{minipage}{\textwidth}
\copyright2018 IEEE. Personal use of this material is permitted. Permission from IEEE must be obtained for all other uses, in any current or future media, including reprinting/republishing this material for advertising or promotional purposes, creating new collective works, for resale or redistribution to servers or lists, or reuse of any copyrighted component of this work in other works.
\end{minipage}}
\maketitle

\begin{abstract}
While becoming more and more present in our every day lives, services that operate on users' locations or location trajectories suffer from general fear of misappropriation of the transmitted location data. Several works have investigated of how to cope with this drawback. Respective systems claim location-privacy, i.e. keeping users' locations secret, by employing anonymisation techniques concerning a user's identity, or by obfuscating the transmitted location. These approaches lead to a degrade of quality-of-service and can be vulnerable to de-anonymisation attacks, or allow to learn at least the approximate location of a user. Focusing on the application domain of \textit{geo-fencing}, we present as remedy a protocol that is based on homomorphic encryption of a user's location. The protocol provably provides full location-privacy by non-exposure of the users' location data, while producing exact geo-fencing results. We provide a detailed definition of the protocol, show its applicability in an actual geo-fencing application, and show that the resulting system fulfills all security properties we see for a location-privacy preserving system.

\end{abstract}


\IEEEpeerreviewmaketitle
\input{Sections/introduction}
\input{Sections/related_work}

\input{Sections/system_model}
\input{Sections/protocol}

\input{Sections/results}

\input{Sections/conclusion}
\input{Sections/acknowledgement}

\bibliographystyle{IEEEtran}
\bibliography{merged_bibliography}

\end{document}

%% file: Sections/introduction.tex
\section{Introduction}
\label{section:introduction}
Location-based services \cite{virrantaus_developing_2001, schiller_location-based_2004} are information services accessible with mobile devices through the mobile network.
They utilise the ability to access, store and analyse real-time geographic information from mobile devices, and, in turn, provide service offerings such as orientation and localisation, navigation, search, identification or checking \cite{reichenbacher_mobile_2004}.

Despite considerable user uptake, e.g., in real-time social systems \cite{cheng_exploring_2011,lindqvist_im_2011} or travel-related applications \cite{yu_personalized_2009,pedrana_location-based_2014}, and promising potentials in commerce \cite{heinemann_study:_2015} or smart cities \cite{singh_location_2015,wang_location_2016}, widespread adoption of location-based services is impeded by 
specific concerns regarding location privacy \cite{beresford2003location} and general fear of misappropriation \cite{dobson_geoslavery_2003}.

These qualms hold in particular for location-tracking services~\cite{barkhuus_location-based_2003}, i.e. location-based services that capture or predict users' location trajectories.
One prime example of location-tracking services is geo-fencing, a class of location-based services that trigger actions or fire events whenever mobile devices enter or exit virtual perimeters set up for geographical areas known as geo-fences.


Clearly, a straight-forward approach to avoid misappropriation of users' location data, are client-based approaches that perform all processing of location information locally on the mobile device. But not only can entirely client-based approaches come with several drawbacks \cite{duckham_location_2006}, they moreover require mobile devices to constantly perform computations against geo-fences locally, using processing power and draining battery of the device.

This paper is therefore concerned with designing a  protocol for network- and cloud-based privacy-preserving geo-fencing services that provably does not leak any location information to any party that is involved with the geo-fencing evaluation. 

Towards this end, we present \textit{NEXUS} (Non-Exposure User location privacy System), a protocol for multi-party geo-fencing evaluation that employs an asymmetric, homomorphic encryption scheme to satisfy non-exposure of the users location, non-exposure of the geo-fences, and computational correctness of the geo-fencing evaluation. The presented approach has been prototypically implemented, being efficient enough to be employed in actual geo-fencing applications.

The remainder of this paper is organised as follows:
Section \ref{sec:related_work} discusses the related literature on privacy preservation in location-based services.
System assumptions, an adversarial model, as well as properties that a location-privacy preserving system should fulfill, are described in Section~\ref{sec:system_model}.
Our NEXUS protocol for location privacy-preserving geo-fencing evaluation is presented in Section~\ref{sec:protocol}, along with an algorithm that, based on the NEXUS protocol, correctly evaluates rectangular-shaped geo-fences; the fulfillment of the previously stated location-privacy properties is discussed subsequently.
We conclude with summary and future work in Section \ref{section:conclusion}.

%% file: Sections/related_work.tex
\section{Related Work}
\label{sec:related_work}
The available literature on location privacy preservation techniques is vast, and can be categorised into four working principles.
We briefly outline the underlying key ideas and refer the interested reader to representative works in each category.

\textbf{Regulation}: Legal frameworks~\cite{ackerman_wireless_2003} governing collection, processing and distribution of individuals' location information have been established in most nations.
However, regulation itself cannot prevent invasion of privacy, often lags behind new technology and innovations, and might stifle location-aware applications.

\textbf{Policy-based approaches}~\cite{cranor_web_2002,bellis_ietf_2014} address the definition of trust-based mechanisms for proscribing certain uses of location information.
These mechanisms are often highly complex and of questionable practicality for highly dynamic location-aware environments.
More importantly, policy systems rely on extratechnological pressures to ensure privacy policies are adhered to. 

\textbf{Anonymisation} approaches rely on the notation of $k$-anonymity, i.e. a node is made indistinguishable from at least $k-1$ other nodes, to prevent location privacy invasion.
Popular techniques for achieving $k$-anonymity are location cloaking~\cite{gruteser_anonymous_2003,chow_peer--peer_2006}, controlled flooding~\cite{ghinita_private_2008} or obfuscation~\cite{duckham_formal_2005}, to name a few. Pseudonymisation~\cite{rodden_lightweight_2002}, a variation of anonymisation, assigns persistent but non-identifying pseudonyms to individuals.
Anonymisation is by far the most popular approach to (location) privacy preservation, however, it presents a barrier to authentication and personalisation, deliberately degrades quality-of-service, and exhibits vulnerabilities to de-anonymisation~\cite{gambs_-anonymization_2013}.

\textbf{Secure multi-party computations} (SMPC) enables parties to jointly compute an arbitrary agreed function of their private inputs with the computation results guaranteed to be correct. 
Prior works investigate multi-party computational geometry~\cite{atallah_secure_2001,li_secure_2005,luo_secure_2007}, privacy-preserving
proximity detection in a two-party setting~\cite{zhong_louis_2007}, privacy-preserving algorithms for determining fair multi-party rendez-vouz points~\cite{bilogrevic_privacy-preserving_2014} or nearest-neighbor queries~\cite{ghinita_private_2008}.
Although closest to our approach, we are not aware of secure multi-party protocols for privacy-preserving geo-fencing.

%% file: Sections/system_model.tex
\section{System, Adversary and Requirements}
\label{sec:system_model}
In the following, we model a geo-fencing system and a privacy-invading adversary.
Ensuing, the privacy properties that our solution should satisfy are outlined.\\
\\
\textbf{System Model.}
We assume the existence of a set of mobile nodes $\mathcal{M}$, a geo-fencing service $\mathcal{G}$, and a set of nodes $\mathcal{S}$ subscribed to geo-fencing events generated by $\mathcal{G}$. 

Each node $M_i \in \mathcal{M}$ may periodically publish its location information in the form $(M_i, \mathcal{E}(\mathbf{L}))$ to the geo-fencing service $\mathcal{G}$, 
where $\mathbf{L} \in \mathbb{R}^2$ specifies $M_i$'s current location and $\mathcal{E}(\mathbf{L})$ denotes a representation of $\mathbf{L}$ suitable for our concerns.

The service $\mathcal{G}$ manages a number of geo-fences, each of which is a tuple of the form $(\mathbf{F},  \mathcal{S}' \subseteq S)$.
We denote a geo-fence's boundary with $\mathbf{F}$, and restrict ourselves to rectangles on the $\mathbb{R}^2$ plane, thus $\mathbf{F} = \{A,B,C,D\}$.
The set $\mathcal{S'}$ contains the nodes subscribed to the geo-fencing events for $(\mathbf{F},  \mathcal{S}' \subseteq S)$ as generated by $\mathcal{G}$.

Upon receiving a tuple $(M_i,\mathcal{E}(\mathbf{L}))$, $\mathcal{G}$ determines if $M_i$ is inside or outside of any eligible geo-fence $(\mathbf{F},  \mathcal{S}')$ by evaluating
\begin{equation}
\label{eqn:fence}
f(\mathcal{E}(\mathbf{L}), \mathbf{F}) = \begin{cases}
1, & \mathbf{L} \in \mathbf{F} \\
0, & \mathbf{L} \notin \mathbf{F}
\end{cases}
\end{equation}
using solely $\mathcal{E}(\mathbf{L})$ and $\mathbf{F}$.

Based on $f(\mathcal{E}(\mathbf{L}), \mathbf{F})$, $\mathcal{G}$ issues notifications to the subscribers in $\mathcal{S}'$. 
Hence, a subscriber $\mathbf{s} \in \mathcal{S}'$ is informed about whether or not a mobile node $M_i \in \mathcal{M}$ is within the geo-fence to which $\mathbf{s}$ was subscribed, however, $\mathbf{s}$ has no knowledge about the particular geo-fence itself.\\
\\
\textbf{Adversary Model.}
 Attacking the system as modeled above, the adversary primarily intends to break a mobile node's location privacy.
That is, the adversary will try to systematically and secretly record any $M_i$'s current or past location for later use.

We assume that mobile nodes in $\mathcal{M}$ publish their locations correctly, and that they cannot be physically tracked by the adversary.
Secondly, we assume $\mathcal{G}$ and nodes in $\mathcal{S}$ to be semi-honest, i.e. these nodes run the protocol exactly as specified but may try to learn as much as possible about the locations of nodes in $\mathcal{M}$.
Finally, we consider the adversary to be computationally bounded.\\
\\
\textbf{Privacy Properties.}
The properties that we intend to provide for a geo-fencing system in the presence of privacy-invading adversaries are given as follows:\\
\begin{itemize}
\item[(P1)] \textbf{Location Non-Exposition}. No party must be must able to obtain $M_i$'s current or past location from interaction or observation.
\item[(P2)] \textbf{Location-privacy Preservation.} No party must be able to learn or deduce $M_i$'s current or past location from interaction or observation.
\item[(P3)] \textbf{Computational Correctness}. Semi-honest parties are guaranteed to produce the correct outputs.
\item[(P4)] \textbf{Network Assistedness.} Since completely client-oriented approaches present several drawbacks~\cite{duckham_location_2006}, \textit{some} information about a mobile nodes' locations must be published to a remote party to not leave all crucial computations to the clients (in our case, the mobile nodes).\\
\end{itemize}

To the best of our knowledge, no previous study has investigated properties (P1) - (P4) in our context.
As outlined before, regulation and policy-based systems rely on extratechnological pressures to ensure privacy;
anonymisation-based approaches certainly violate (P1) and (P3) and exhibit vulnerabilities with respect to (P2);
SMPC approaches do not yet address geo-fencing services at all.

%% file: Sections/protocol.tex
\section{Protocol Specification and Implementation}
\label{sec:protocol}
Obviously, encryption seems to be a suitable way to represent $\mathbf{L}$ to $\mathcal{G}$. The chosen encryption scheme needs to provide certain characteristics to allow evaluation of $f$ solely based on $\mathcal{E}(\mathbf{L})$.

For this, in the following, we first provide some basics about cryptography and homomorphic encryption.
Ensuing, we introduce our Non-Exposure User location privacy System (NEXUS) protocol for geo-fencing services, based on a homomorphic public key encryption scheme.
Finally, we provide an overview of our prototype implementation and detail on how to perform geo-fence containment tests within the constraints of our chosen encryption system.\\
\\
\textbf{Preliminaries.}
A conventional encryption scheme consists of the following functions:
\begin{itemize}
\item \textbf{Key generation function} generates a symmetric encryption and decryption key $ek$ or an asymmetric~\cite{rivest1978method} public-private key pair $(pk,sk)$ based on some parameter.
\item \textbf{Encryption function} $\mathcal{E}_k(x)$ outputs a ciphertext from plaintext $x$ using a key $k$.
\item \textbf{Decryption function} $\mathcal{D}_k(x)$ outputs a plaintext from ciphertext $x$ using a key $k$.
\end{itemize}
For any plaintext $x$, one has $\mathcal{D}_{ek}(\mathcal{E}_{ek}(x)) = x$ for symmetric schemes and $\mathcal{D}_{sk}(\mathcal{E}_{pk}(x)) = x$ for asymmetric encryption schemes.

In addition, \textit{homomorphic encryption}~\cite{RAD1978,FG2007} features characteristics
\begin{equation}
\label{eq:homomorphic_encryption}
\small
\mathcal{E}_k(m_1) \odot \mathcal{E}_k(m_2) = \mathcal{E}_k(m_1 \bullet m_2)
\end{equation}
In other words, a homomorphic encryption scheme enables specific computations on ciphertexts and generates encrypted results which, when decrypted, match the results of operations performed on the respective plaintexts.

\begin{figure}[!t]
\centering
\includegraphics[width=2.3in]{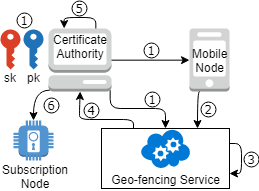}
\caption{The actors and high level interactions in NEXUS.}
\label{fig:actors}
\end{figure}

We will in the following show that homomorphic encryption is thus suitable to evaluate $f$ in Equation (\ref{eqn:fence}) solely by the provided $\mathcal{E}_k(\mathbf{L})$, as required. As homomorphic encryption requires to encrypt all operands with the same key (cf. Equation (\ref{eq:homomorphic_encryption})), the chosen encryption scheme must be a public key encryption scheme, as otherwise, parties that need to encrypt values during the process could use the symmetric key to decrypt $\mathcal{E}_{ek}(\mathbf{L})$.
\\ \\
\textbf{Protocol.}
Our protocol is based on a public key infrastructure (PKI) and a suitable homomorphic encryption scheme. 
We do not rely on the geo-fencing service $\mathcal{G}$ acting as a trusted third party (TTP), but rather on the collaboration of $\mathcal{G}$ and a certificate authority and evaluation service $\mathcal{A}$ taking the role to provide network assistedness. Hence, we extend our system model as indicated in Figure~\ref{fig:actors}. The protocol then is as follows:

\ding{192} The certificate authority $\mathcal{A}$ generates a public-private key pair $(pk,sk)$ and shares $pk$ on demand with every other party.

\ding{193} Each $M_i$ may periodically send information about its location $\mathbf{L}$ in the form $(M_i, \mathcal{E}_{pk}(\mathbf{L}))$ to $\mathcal{G}$.

\ding{194} $\mathcal{G}$ can not decrypt $\mathcal{E}_{pk}(\mathbf{L})$ or any results of homomorphic computations based on $\mathcal{E}_{pk}(\mathbf{L})$, but instead homorphically computes an (encrypted) intermediate result $\mathbf{R}$ as
\begin{equation}
\label{eq:fg_def}
\mathbf{R} = f_\mathcal{G}(\mathcal{E}_{pk}(\mathbf{L}), \mathbf{F})
\end{equation}
with $f_\mathcal{G}$ being a function that homomorphically operates on $\mathcal{E}_{pk}(\mathbf{L})$, and $\mathbf{F}$ the boundary of a geo-fence $(\mathbf{F}, \mathcal{S'})$ as defined in Section \ref{sec:system_model}. $\mathcal{G}$ can encrypt $\mathbf{F}$ with $pk$ for homomorphic operations, where necessary.

\ding{195} $\mathcal{G}$ passes $\mathbf{R}$, the ID of the mobile node $M_i$, and the set of subscription nodes $\mathcal{S'}$ that are to be invoked based on the evaluation result to $\mathcal{A}$.

\ding{196} $\mathcal{A}$ then evaluates $\mathbf{L} \in \mathbf{F}$ by an evaluation function $f_\mathcal{A}$, such that
\begin{equation*}
\mathbf{L} \in \mathbf{F} \Leftrightarrow f_\mathcal{A}(\mathbf{R}) =  \mathtt{true}
\end{equation*}

$\mathcal{A}$ can decrypt $\mathbf{R}$ with $sk$, where necessary. By this, $f$ as given in Equation (\ref{eqn:fence}) is computed as
\begin{equation*}
f(\mathcal{E}_{pk}(\mathbf{L}), \mathbf{F}) = [f_\mathcal{A} \circ f_\mathcal{G}] (\mathcal{E}_{pk}(\mathbf{L}), \mathbf{F})
\end{equation*}
\ding{197} $\mathcal{A}$ will invoke the respective subscriber $\mathbf{s} \in \mathcal{S'}$ with the result of the evaluation and the ID of the mobile node $M_i$.
\\ \\
\textbf{Implementation.}
We base our implementation of above protocol on the \textit{Paillier} encryption scheme. Implementation details concerning key generation, encryption, and decryption functions are found in~\cite{P1999}.
Paillier features homomorphic addition:
\begin{equation}
\label{eq:paillier_add}
\mathcal{E}_k(m_1) \oplus \mathcal{E}_k(m_2) = \mathcal{E}_k(m_1) * \mathcal{E}_k(m_2) = \mathcal{E}_k(m_1 + m_2) , \\
\end{equation}

and pseudo-homomorphic multiplication of a ciphertext by an unsigned integer $u$:

\begin{equation}
\label{eq:paillier_mul}
\mathcal{E}_k(m)^u = \mathcal{E}_k(u * m),
\end{equation}

with $+$ and $*$ being arithmetic addition and multiplication.
Moreover, using Equations (\ref{eq:paillier_add}) and (\ref{eq:paillier_mul}), we can homomorphically subtract by adding a negated ciphertext.
\begin{equation}
\label{eq:paillier_sub}
\small
\mathcal{E}_k(m_1) \ominus \mathcal{E}_k(m_2) = \mathcal{E}_k(m_1) \oplus \mathcal{E}_k(m_2)^{-1} = \mathcal{E}_k(m_1-m_2)
\end{equation}

\begin{figure}[!t]
\centering
\includegraphics[width=1.6in]{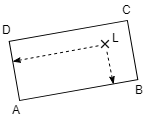}
\caption{To check if a point is inside a rectangle, the point is projected perpendicular to the sides of the rectangle}
\label{fig:rect}
\end{figure}

In our implementation, every position is a two dimensional vector consisting of a longitude and latitude value, which we indicate by indices $lon$ and $lat$.
We determine if a position $\mathbf{L}$ is inside a rectangle by testing perpendicular projection (see Figure \ref{fig:rect}) as follows:
\begin{equation}
\label{eq:evaluation_term}
\small
0 \leq \vv{A\mathbf{L}} \cdot \vv{AB} \leq \vv{AB}^2 \wedge 0 \leq \vv{A\mathbf{L}} \cdot \vv{AD} \leq \vv{AD}^2
\end{equation}
Upon receiving a tuple $(M_i, \mathcal{E}_{pk}(\mathbf{L})$), $\mathcal{G}$ computes for a rectangular geo-fence $\mathbf{F} = \{A, B, C, D\}$ the terms of Equation (\ref{eq:evaluation_term}) as
\begin{equation*}
\small
\mathbf{R} = \{\vv{AB}^2, \vv{AD}^2, \mathcal{E}_{pk}(\vv{A\mathbf{L}} \cdot \vv{AB}), \mathcal{E}_{pk}(\vv{A\mathbf{L}} \cdot \vv{AD})\},
\end{equation*}
constituting $f_\mathcal{G}$, cf. Equation (\ref{eq:fg_def}).
\\

$\mathcal{G}$ has to calculate $\mathcal{E}_{pk}(\vv{A\mathbf{L}} \cdot \vv{AB})$ and $\mathcal{E}_{pk}(\vv{A\mathbf{L}} \cdot \vv{AD})$ homomorphically due to encrypted $\mathbf{L}$. For this, $\mathcal{G}$ encrypts $\mathbf{F}$ with the shared public key $pk$. One can show that with Equations (\ref{eq:paillier_add}), (\ref{eq:paillier_mul}) and (\ref{eq:paillier_sub}), definition of dot product, and component-wise computation, it holds that

\begin{equation*}
\label{eq:component_wise}
\mathcal{E}_{pk}(\vv{A\mathbf{L}}_{lat}) = \mathcal{E}_{pk}(\mathbf{L}_{lat}) \ominus \mathcal{E}_{pk}(A_{lat}),
\end{equation*}
similar for longitude components, and with that
\begin{equation*}
\label{eq:hom_check}
\small
\mathcal{E}_{pk}(\vv{A\mathbf{L}}\cdot\vv{AB}) = \mathcal{E}_{pk}(\vv{A\mathbf{L}}_{lat})^{\vv{AB}_{lat}} \oplus \mathcal{E}_{pk}(\vv{A\mathbf{L}}_{lon})^{\vv{AB}_{lon}},
\end{equation*}
and similar for $\mathcal{E}_{pk}(\vv{A\mathbf{L}} \cdot \vv{AD})$.

The calculated values $\mathbf{R}$ are then sent to the certificate authority and evaluation service $\mathcal{A}$ along with the node ID $M_i$ and the subscription nodes $\mathcal{S}'$ of the actions to trigger. $\mathcal{A}$ decrypts the encrypted results with the secret key $sk$, evaluates Equation (\ref{eq:evaluation_term}) as $f_\mathcal{A}$, and invokes the according subscribers.
\\ \\
\textbf{Performance evaluation}. For our performance evaluation deployment, we implemented $\mathcal{G}$ and $\mathcal{A}$ as Python applications with Flask\footnote{\url{http://flask.pocoo.org/}} as backend Web-server. Same holds for the prototype client that we used for testing. Parties communicate with HTTP requests.
For the homomorphic cryptography, we used the n1analytics/python-paillier Python library\footnote{\url{https://github.com/n1analytics/python-paillier}}.
On a workstation (Linux, i5-2520M at 2.50GHz, 16GB RAM), we measured evaluation times of an average 323 milliseconds (1000 executions, network lag not included) per evaluation of $f_\mathcal{A} \circ f_\mathcal{G}$ with a 2048-bit Paillier key. The measured time is in the ranges suitable for an execution that is perceived ''uninterrupted'' by users~\cite{card1991information}. Thus, we consider NEXUS practicable for actual application.

%% file: Sections/results.tex
\section{Discussion and Results}
\label{sec:results}

The goal must be, that by our architecture and protocol, the security properties (P1) to (P4) can never be violated by a semi-honest adversary.

Adversaries, in our presented architecture the geo-fencing service $\mathcal{G}$, certificate authority and evaluation service $\mathcal{A}$, and subscriptions $\mathcal{S}$, could try to obtain $M_i$'s secret location $\mathbf{L}$ directly (violating (P1)), which would require them to be in possession of both the encrypted location $\mathcal{E}_{pk}(\mathbf{L})$ and the secret key $sk$ to decrypt $L$.

Or they could try to approximate $\mathbf{L}$ by observing evaluation of geo-fences $(\mathbf{F}, \mathcal{S'})$. For this, they would either need to learn about both $\mathbf{F}$ and the outcome of $f(\mathcal{E}_{pk}(\mathbf{L}), \mathbf{F})$ (Equation (\ref{eqn:fence})). Or they need to know about the set $\mathcal{S' \subseteq S}$ that is registered to a geo-fence $(\mathbf{F}, \mathcal{S'})$. Observing then if any $\mathbf{s} \in \mathcal{S'}$ was invoked as a result of an evaluation would allow to deduct the evaluation outcome against a geo-fence boundary $\mathbf{F}$.

In the following, we show that by employing the protocol as described in Section \ref{sec:protocol}, none of these attacks are possible, and by this, the security properties (P1) to (P4) as stated in Section \ref{sec:system_model} are always fulfilled:
\\ \\
\textbf{Location Non-Exposition (P1)}: The geo-fencing service $\mathcal{G}$ only receives the encrypted location $\mathcal{E}_{pk}(\mathbf{L})$ from a mobile node $M_i$. $\mathcal{G}$ is not in possession of the private key $sk$ to decrypt the data, nor does it need to decrypt any value by exploiting the homomorphic characteristics of the Paillier system as described in \ref{sec:protocol}. $\mathcal{G}$  can thus at no point obtain $\mathbf{L}$ directly.
 The Authority service $\mathcal{A}$, though in possession of the private key $sk$, never receives any encrypted location $\mathcal{E}_{pk}(\mathbf{L})$ from neither $M_i$ nor $\mathcal{G}$. Thus, $\mathcal{A}$ does at no point obtain information about $\mathbf{L}$. Subscribers in $S$ receive neither $sk$ nor the encrypted location $\mathcal{E}_{pk}(\mathbf{L})$. (P1) is by this always satisfied.
\\ \\
\textbf{Location-privacy Preservation (P2)}: With $\mathcal{G}$ defining the geo-fences $(\mathbf{F}, \mathcal{S'})$, it is clearly in knowledge about both $\mathbf{F}$ and $\mathcal{S'}$. $\mathcal{G}$ is not in possession of $sk$ and by this can not decrypt the evaluation result and invoke any of the subscriptions $\mathcal{S'}$ based on it. $\mathcal{G}$ can thus neither obtain, nor observe the result of an evaluation for a geo-fence $(\mathbf{F}, \mathcal{S'})$.

	$\mathcal{G}$ does not disclose information about geo-fence boundaries $\mathbf{F}$ to any party,
and in particular no information about to which geo-fence $(\mathbf{F}, \mathcal{S'})$  subscriptions $\mathbf{s} \in \mathcal{S'}$ are registered. $\mathcal{A}$ by this can never observe for which geo-fence $(\mathbf{F}, \mathcal{S'})$ it decrypted the evaluation result.
To retrieve $\mathbf{L}$ from the decrypted values $\vv{A\mathbf{L}} \cdot \vv{AB}$ and $\vv{A\mathbf{L}} \cdot \vv{AD}$, $\mathcal{A}$ would need knowledge about the geo-fence boundary $\mathbf{F} = \{A, B, C, D\}$, which is never disclosed by $\mathcal{G}$.

	Subscribers in $\mathbf{s} \in S$ receive the result of an evaluation, but do not have knowledge about the geo-fence boundary $\mathbf{F}$, as $\mathbf{F}$ is not disclosed by $\mathcal{G}$. By this, a subscriber can not observe the location of a mobile node by observing its own invocation.

None of the transmitted data or computational intermediate results are stored during the process. By this, (P2) is always fulfilled.
\\ \\
\textbf{Computational Correctness (P3)}: None of the parties willingly provides wrong data during the process, as we assume all participants to be semi-honest. Correctness of computations during the evaluation is moreover ensured by the homomorphic characteristics of the chosen Paillier encryption scheme. We have shown the correct computation of a containment check of $\mathbf{L}$ against a rectangle based on the Paillier homomorphism in Section \ref{sec:protocol}. (P3) is by this always fulfilled.
\\ \\
\textbf{Network Assistedness (P4)}: Mobile nodes do not receive geo-fences from $\mathcal{G}$ and by this can not perform geo-fencing evaluation computations on their own.  They need to provide \textit{some} information to a third party, in this case, $\mathcal{E}_{pk}(\mathbf{L})$ to $\mathcal{G}$, which performs the evaluation with the help of $\mathcal{A}$. As shown above, information exchanged during this process is not sufficient for any other party to violate (P1) to (P3), and (P4) is always fulfilled.

%% file: Sections/conclusion.tex
\section{Future Work}
\label{section:future_work}
We have in this paper limited ourselves to rectangular geo-fence shapes. For future work, we plan to investigate how to evaluate containment of a users' position in arbitrarily shaped geo-fences within the possible mathematical operations as imposed by the encryption schemes.

Geo-fences are defined independently. This allows for parallel evaluation of multiple geo-fences. We plan to do further research on highly scalable large-scale distributed setups of several instances of geo-fencing and evaluation services, employing capabilities of latest Infrastructure-as-a-Service and virtual container management systems.

We moreover see in the proposed solution a promising approach to realise general purpose validation tasks in distributed IoT environments, similar to works on homomorphic privacy preserving multi party computations as for example presented in \cite{cramer2001multiparty}. In fact, while we focused on the geo location validation as use-case in this paper, we see promising application opportunities for every kind of numerical computations or validations, like for example factory process sanitary monitoring in an Industrie 4.0 scenario, general sensor analysis in a Smart City environment, and other.

\section{Conclusion}
\label{section:conclusion}

In this paper, we have presented NEXUS (Non Exposure User location privacy System), a novel protocol for geo-fencing systems that ensures location privacy based on non-location exposure by employing a homomorphic encryption scheme.

In the paper, we have presented the following contributions to the topic:

  Based on a list of properties for location-privacy preserving geo-fencing schemes, we have developed a protocol for a geo-fencing system that, unlike existing approaches, does not rely on  anonymisation or obfuscation of the user. Instead, it utilises characteristics of homomorphic encryption. This allows to perform the evaluation entirely on encrypted location data. We have shown that by our protocol, provably, the correct result of the geo-fencing evaluation can be computed. The location of the client is kept entirely secret to all parties involved with the evaluation process.

  The presented protocol was implemented in a prototype application for rectangular geo-fences based on the Paillier encryption system \cite{P1999}. The prototype performed the computation in time ranges that allow for an actual applicable system.

%% file: Sections/acknowledgement.tex
\section*{Acknowledgement}
The work in this paper was funded by financial means of the German Federal Ministry for Economic Affairs and Energy (BMWi) in the project GUIDED AL under the support code 01MD16010C, and by the Federal Ministry of Education and Research of Germany in the project Hybr-iT under  support code 01IS16026A.